\documentclass[12pt]{iopart}
\usepackage[utf8]{inputenc}
\usepackage[T1]{fontenc}
\expandafter\let\csname equation*\endcsname=\relax
\expandafter\let\csname endequation*\endcsname=\relax
\usepackage{multicol}
\usepackage[intlimits]{amsmath}
\usepackage{amssymb,amsfonts,amsthm}
\usepackage{xspace,graphicx}

\theoremstyle{definition}

\theoremstyle{remark}

\catcode`,\active

\catcode`\,12

\catcode`,\active

\catcode`\,12

\newcommand{\pd}{\partial}

\newcommand{\braket}[2]{\ensuremath{\langle #1|#2\rangle}\xspace}
\newcommand{\under}[1]{_{#1}}
\numberwithin{equation}{section}

\begin{document}
\title[The singular and the $2:1$ anisotropic Dunkl oscillators]{The singular and the $2:1$ anisotropic Dunkl\\ oscillators in the plane}
\author{Vincent X. Genest}
\ead{genestvi@crm.umontreal.ca}
\address{Centre de Recherches Math\'ematiques, Universit\'e de Montr\'eal, C.P. 6128, Succursale Centre-ville, Montr\'eal, Qu\'ebec, Canada, H3C 3J7}
\author{Luc Vinet}
\ead{luc.vinet@umontreal.ca}
\address{Centre de Recherches Math\'ematiques, Universit\'e de Montr\'eal, C.P. 6128, Succursale Centre-ville, Montr\'eal, Qu\'ebec, Canada, H3C 3J7}
\author{Alexei Zhedanov}
\ead{zhedanov@yahoo.com}
\address{Donetsk Institute for Physics and Technology, Donetsk 83114, Ukraine}
\begin{abstract}
Two Dunkl oscillator models are considered: one singular and the other with a $2:1$ frequency ratio. These models are defined by Hamiltonians which include the reflection operators in the two variables $x$ and $y$. The singular or caged Dunkl oscillator is second-order superintegrable and  admits separation of variables in both Cartesian and polar coordinates. The spectrum of the Hamiltonian is obtained algebraically and the separated wavefunctions are given in the terms of Jacobi, Laguerre and generalized Hermite polynomials. The symmetry generators are constructed from the $\mathfrak{su}(1,1)$ dynamical operators of the one-dimensional model and generate a cubic symmetry algebra. In terms of the symmetries responsible for the separation of variables, the symmetry algebra of the singular Dunkl oscillator is quadratic and can be identified with a special case of the Askey-Wilson algebra $AW(3)$ with central involutions. The $2:1$ anisotropic Dunkl oscillator model is also second-order superintegrable. The energies of the system are obtained algebraically, the symmetry generators are constructed using the dynamical operators and the resulting symmetry algebra is quadratic. The general system appears to admit separation of variables only in Cartesian coordinates. Special cases where separation occurs in both Cartesian and parabolic coordinates are considered. In the latter case the wavefunctions satisfy the biconfluent Heun equation and depend on a transcendental separation constant.
\end{abstract}
\section{Introduction}
This paper purports to analyze the singular and the $2:1$ anisotropic Dunkl oscillator models in the plane. These two-dimensional quantum systems are defined by Hamiltonians of Dunkl type which involve the reflection operators in the $x$ and $y$ variables. As shall be seen, these two models exhibit many interesting properties: they are second-order superintegrable, exactly solvable and, in certain cases, they allow separation of variables in more than one coordinate system.

A quantum system with $n$ degrees of freedom described by a Hamiltonian $H$ is (maximally) superintegrable if it possesses $2n-1$ algebraically independent symmetry generators $S_i$ such that
\begin{align*}
[H,S_i]=0,\qquad i=1,\ldots,2n-1,
\end{align*}
where one of the symmetries is the Hamiltonian itself. For such a system, it is impossible for all the symmetry generators to commute with one another and hence the $S_i$ generate a non-Abelian symmetry algebra.  If $m$ is the maximal order of the symmetry operators (apart from $H$) in the momenta, the system is said to be $m$\textsuperscript{th}- order superintegrable. 

First order superintegrability is associated to geometrical symmetries and to Lie algebras \cite{Olver-1993} whereas second order superintegrability is typically associated to quadratic symmetry algebras \cite{Zhedanov-1992-2,Zhedanov-1992,Granovskii-1992,Kalnins-1996,Post-2008} and to separation of variables in more than one coordinate system \cite{Winter-1967-2,Winter-1965,Kalnins-2005-I,Kalnins-2005-V,Miller-1977}. For example, in the Euclidean plane, all second-order superintegrable systems of the general form 
\begin{align*}
H=-\frac{1}{2}\nabla^2+V(x,y),
\end{align*}
are known and have been classified \cite{Winter-1967}. The possible systems are the singular or caged oscillator:
\begin{align}
\label{Standard-Singular}
V(x,y)=\omega(x^2+y^2)+\frac{\alpha}{x^2}+\frac{\beta}{y^2},
\end{align}
which separates in Cartesian and polar coordinates; the anisotropic oscillator with a $2:1$ frequency ratio:
\begin{align}
\label{Standard-Anisotropic}
V(x,y)=\omega(4x^2+y^2)+\frac{\gamma}{y^2},
\end{align}
which separates in Cartesian and parabolic coordinates and the Coulomb problem:
\begin{align*}
V(r,\phi)=\frac{\alpha}{2r}+\frac{1}{4r^2}\left(\frac{\beta_1}{\cos^2(\phi/2)}+\frac{\beta_2}{\sin^2(\phi/2)}\right),
\end{align*}
which separates in polar and parabolic coordinates. The fourth superintegrable system admits separation in two mutually perpendicular parabolic coordinate systems. We note in passing that only the first two systems \eqref{Standard-Singular} and \eqref{Standard-Anisotropic} are genuinely different by virtue of the Levi-Civita mapping \cite{Vinet-1995}; this topic is discussed in the conclusion.

In view of the special properties and applications of superintegrable models, there is considerable interest in enlarging the set of documented systems with this property. Recent advances in this perspective include the study of superintegrable systems with higher order symmetries \cite{Miller-2011,Miller-2012,Marquette-2010,TTW-2009, TTW-2010}, the construction of new superintegrable models from exceptional polynomials \cite{Marquette-2013-1,Post-2012-3}, the search for discretized superintegrable systems \cite{Post-2012} and the examination of models described by Hamiltonians involving reflection operators \cite{Genest-2013-1,Genest-2013-2,VDJ-2011,VDJ-2012,Plyu-1996,Plyu-1997,Post-2012-2,Post-2011-3}.

Hamiltonians that include reflection operators have most notably occurred in the study of integrable systems of Calogero-Sutherland type \cite{Calogero-1969,Lapointe-1996-2,Lapointe-1996,Sutherland-1971} and their generalizations \cite{Minahan-1993,Brink-1993}. They also arise in the study of parabosonic oscillators \cite{Mukunda-1980,Rozen-1994,Mukunda-1981}. These models are best described in terms of Dunkl operators \cite{Dunkl-2000}, which are differential/difference operators that include reflections \cite{Dunkl-1989}. These operators are central in the theory of multivariate orthogonal polynomials \cite{Dunkl-2001} and are at the heart of Dunkl harmonic analysis \cite{Rosler-1998}, which is currently under active development . Furthermore, the recent study of polynomial eigenfunctions of first and second order differential/difference operators of Dunkl type has led to the discovery of several new families of classical orthogonal polynomials of a single variable known as $-1$ polynomials, also referred to as polynomials of Bannai-Ito type \cite{Genest-2012-1,Vinet-2012-1,Vinet-2012-2,Vinet-2011,Vinet-2012-4,Vinet-2012-3}. These new polynomials are related to Jordan algebras \cite{Genest-2012-3,Tsujimoto-2011} and quadratic algebras with reflections \cite{Genest-2012-1,Genest-2012-2,Zhedanov-2011}.

This motivates the study of superintegrable and exactly solvable models that involve reflections. Recently, we introduced the Dunkl oscillator model in the plane \cite{Genest-2013-1,Genest-2013-2} described by the Hamiltonian
\begin{align}
\label{Dunkl-Osc}
\mathcal{H}=-\frac{1}{2}[(\mathcal{D}_{x}^{\mu_x})^2+(\mathcal{D}_{y}^{\mu_y})^2]+\frac{1}{2}(x^2+y^2),
\end{align}
where $\mathcal{D}_{x}^{\mu_x}$ stands for the Dunkl derivative
\begin{align}
\label{Dunkl-D}
\mathcal{D}_{x}^{\mu_x}=\pd_{x}+\frac{\mu_x}{x}(1-R_x),
\end{align}
where $R_xf(x)=f(-x)$ is the reflection operator. This is possibly the simplest two-dimensional model with reflections and it corresponds to the combination of two independent parabosonic oscillators \cite{Rozen-1994}. The Dunkl oscillator has been shown to be second-order superintegrable and its wavefunctions, overlap coefficients and symmetry algebra have been related to $-1$ polynomials.

We shall here consider two extensions of the Hamiltonian \eqref{Dunkl-Osc}. The first one, called the singular Dunkl oscillator, corresponds to the Hamiltonian \eqref{Dunkl-Osc} with additional singular terms proportional to $x^{-2}$ and $y^{-2}$. The second one, called the $2:1$ anisotropic Dunkl oscillator, corresponds to a singular Dunkl oscillator in the $y$ direction combined with a Dunkl oscillator with twice the frequency in the $x$ direction. 

The two-dimensional singular Dunkl oscillator will be shown to be second-order superintegrable and to admit separation of variables in Cartesian and polar coordinates. Its separated wavefunctions will be obtained in terms of Jacobi, Laguerre and generalized Hermite polynomials. A cubic symmetry algebra with reflections will be found for this model, as opposed to the linear Lie-type algebra extended with reflections obtained for the ordinary Dunkl oscillator \eqref{Dunkl-Osc} in \cite{Genest-2013-1}. In terms of the symmetries responsible for the separation of variables, the invariance algebra is quadratic and will be identified to the Hahn algebra with central involutions; the Hahn algebra is a special case of the Askey-Wilson algebra $AW(3)$ \cite{Zhedanov-1992-2}. The appearance of the Hahn algebra as symmetry algebra will also establish that the expansion coefficients between the Cartesian and polar bases are given in terms of the dual Hahn polynomials. 

The anisotropic Dunkl oscillator will also be shown to be second-order superintegrable and its quadratic symmetry algebra will be constructed with the dynamical (spectrum-generating) operators of the one-dimensional components. It will be seen that for this model the separation of variables is not possible in general. Special cases where separation in parabolic coordinates do occur will be examined; they correspond to the combination of either a singular or  an ordinary Dunkl oscillator in one direction with a standard harmonic oscillator with twice the frequency in the other direction. We shall show in one of these special cases that the wavefunctions in parabolic coordinates are expressed in terms of biconfluent Heun functions which depend on a transcendental parameter.
 
The organization of the remainder of this article is straightforward. Section 2 is dedicated to the analysis of the singular oscillator. Section 3 bears on the $2:1$ anisotropic Dunkl oscillator. Section 4 concludes the paper with remarks on the Dunkl-Coulomb problem and on the Levi-Civita mapping for models involving Dunkl derivatives.

\section{The singular Dunkl oscillator}
In this section, the singular Dunkl oscillator model in the plane is introduced. The model can be considered both as a generalization of the model \eqref{Standard-Singular} with the standard derivatives replaced by the Dunkl derivatives or as an extension of the Hamiltonian \eqref{Dunkl-Osc} with additional singular terms in the potential.
\subsection{Hamiltonian, dynamical symmetries and spectrum}
The singular Dunkl oscillator in the plane is described by the Hamiltonian 
\begin{equation}
\label{Dunkl-Singular-Osc}
\mathcal{H}=-\frac{1}{2}\Big[(\mathcal{D}_{x}^{\mu_x})^2+(\mathcal{D}_{y}^{\mu_y})^2\Big]+\frac{1}{2}(x^2+y^2)+\frac{(\alpha_{x}+\beta_xR_{x})}{2x^2}+\frac{(\alpha_{y}+\beta_{y}R_{y})}{2y^2},
\end{equation}
where $\mathcal{D}_{x_i}^{\mu_{x_i}}$ is the Dunkl derivative \eqref{Dunkl-D} whose square has the expression
\begin{align*}
(\mathcal{D}_{x_i}^{\mu_{x_i}})^2=\pd_{x_i}^2+\frac{2\mu_{x_i}}{x_i}\pd_{x_i}-\frac{\mu_{x_i}}{x_i^2}(1-R_{x_i}),\qquad \pd_{x_i}=\frac{\pd}{\pd x_i},
\end{align*}
and where $R_{x_{i}}$ is the reflection operator 
\begin{align*}
R_{x_i}f(x_i)=f(-x_i),\qquad i=1,2,
\end{align*}
with $x_1=x$ and $x_2=y$. The parameters $\alpha_{x_i}$, $\beta_{x_i}$ obey the quantization conditions
\begin{subequations}
\label{Parameter-Quantization}
\begin{align}
\alpha_{x_i}&=2k_{x_i}^{+}(k_{x_i}^{+}+\mu_{x_i}-1/2)+2k_{x_i}^{-}(k_{x_i}^{-}+\mu_{x_i}+1/2),\\
\beta_{x_i}&=2k_{x_i}^{+}(k_{x_i}^{+}+\mu_{x_i}-1/2)-2k_{x_i}^{-}(k_{x_i}^{-}+\mu_{x_i}+1/2),
\end{align}
\end{subequations}
with $k_{x_i}^{\pm}\in\mathbb{Z}$. The quantization conditions \eqref{Parameter-Quantization} can be seen to arise from the parity requirements (due to the reflections) on the solutions of the Schr\"odinger equation associated to the Hamiltonian \eqref{Dunkl-Singular-Osc} (see subsection $2.2.1$).

Strikingly, the singular Dunkl oscillator \eqref{Dunkl-Singular-Osc} exhibits a $\mathfrak{su}(1,1)$ dynamical symmetry similar to that of the ordinary singular oscillator \cite{Vinet-1995}. To see this, one first introduces two commuting sets $(a_{x},a_{x}^{\dagger})$, $(a_{y},a_{y}^{\dagger})$ of parabosonic creation/annihilation operators \cite{Mukunda-1980,Rozen-1994} :
\begin{align}
\label{Dunkl-Creation-Annihilation}
a_{x_i}=\frac{1}{\sqrt{2}}\left(x_i+\mathcal{D}_{x_i}^{\mu_{x_i}}\right),\quad a_{x_i}^{\dagger}=\frac{1}{\sqrt{2}}\left(x_i-\mathcal{D}_{x_i}^{\mu_{x_i}}\right).
\end{align}
These operators satisfy the following commutation relations:
\begin{align*}
[a_{x_i},a_{x_i}^{\dagger}]=1+2\mu_{x_i}R_{x_i},\quad \{a_{x_i},R_{x_i}\}=0,\quad \{a_{x_i}^{\dagger},R_{x_i}\}=0,
\end{align*}
where $[a,b]=ab-ba$ and $\{a,b\}=ab+ba$. Upon defining the generators
\begin{align}
\label{Dynamical-OPS}
A_{x_i}^{\dagger}=(a_{x_i}^{\dagger})^2-\frac{(\alpha_{x_i}+\beta_{x_i}R_{x_i})}{2x_i^2},\quad A_{x_i}=(a_{x_i})^2-\frac{(\alpha_{x_i}+\beta_{x_i}R_{x_i})}{2 x_i^2},
\end{align}
a direct computation shows that
\begin{align}
\label{Singular-Dynamical-Algebra}
[\mathcal{H}_{x_i},A_{x_i}^{\dagger}]=2 A_{x_i}^{\dagger},\quad [\mathcal{H}_{x_i},A_{x_i}]=-2 A_{x_i},\quad [A_{x_i}^{\dagger},A_{x_i}]=-4\mathcal{H}_{x_i},
\end{align}
where $\mathcal{H}_{x_i}$ is the Hamiltonian of the one-dimensional singular Dunkl oscillator
\begin{align}
\label{Singular-Dunkl-1D}
\mathcal{H}_{x_i}=-\frac{1}{2}(\mathcal{D}_{x_i}^{\mu_{x_i}})^2+\frac{x_i^2}{2}+\frac{(\alpha_{x_i}+\beta_{x_i}R_{x_i})}{2x_i^2}.
\end{align}
It is also easily verified that
\begin{align}
\label{Full-Commute}
[\mathcal{H}_{x_i},R_{x_i}]=0,\quad [A_{x_i}^{\dagger},R_{x_i}]=0,\quad [A_{x_i},R_{x_i}]=0.
\end{align}
The algebra \eqref{Singular-Dynamical-Algebra} is forthwith identified with the Lie algebra $\mathfrak{su}(1,1)$. Indeed, upon taking
\begin{align}
\label{Identification-Algebra}
2J_{0}=\mathcal{H}_{x_i},\quad 2J_{+}=A_{x_i}^{\dagger},\quad 2J_{-}=A_{x_i},
\end{align}
the defining relations of $\mathfrak{su}(1,1)$ are recovered:
\begin{align}
\label{SU(1,1)}
[J_0,J_{\pm}]=\pm J_{\pm},\quad [J_{+},J_{-}]=-2J_{0}.
\end{align}
We also have in this case $J_{\pm}^{\dagger}=J_{\mp}$. The Casimir operator $\mathcal{C}$ of the algebra \eqref{SU(1,1)} is of the form
$$
\mathcal{C}=J_0^2-J_{+}J_{-}-J_{0}.
$$
In the realization \eqref{Dynamical-OPS}, \eqref{Singular-Dynamical-Algebra}, \eqref{Singular-Dunkl-1D} the Casimir operator may be expressed as
\begin{align*}
C_{x_i}=\mathcal{H}_{x_i}^2-A^{\dagger}_{x_i}A_{x_i}-2\mathcal{H}_{x_i},
\end{align*}
and is seen to have the following action on functions of argument $x_i$:
\begin{align*}
C_{x_i}f(x_i)=(\mu_{x_i}^2+\alpha_{x_i}-3/4)f(x_i)+(\beta_{x_i}-\mu_{x_i})f(-x_i).
\end{align*}
Recall that the reflection operator $R_{x_i}$ commutes with all the generators and can thus be simultaneously diagonalized with $C_{x_i}$. The operator $C_{x_i}$ hence take two possible values depending on the parity of $f(x_i)$. On even functions, one has
\begin{align}
\label{Even}
C_{x_i}f(x_i)=4\,\delta_{x_i}(\delta_{x_i}-1)f(x_i),\qquad \delta_{x_i}=k_{x_i}^{+}+\mu_{x_i}/2+1/4,
\end{align}
and on odd functions, one finds
\begin{align}
\label{Odd}
C_{x_i}f(x_i)=4\,\epsilon_{x_i}(\epsilon_{x_i}-1)f(x_i),\qquad \epsilon_{x_i}=k_{x_i}^{-}+\mu_{x_i}/2+3/4.
\end{align}
It is possible to introduce an invariant operator $Q_{x_i}$ given by
\begin{align}
\label{Dynamical-Casimir}
Q_{x_i}=\mathcal{H}_{x_i}^2-A_{x_i}^{\dagger}A_{x_i}-2\mathcal{H}_{x_i}+(\mu_{x_i}-\beta_{x_i})R_{x_i},
\end{align}
which commutes with all the generators $\mathcal{H}_{x_i}$, $A_{x_i}$, $A_{x_i}^{\dagger}$ and acts as a multiple of the identity on the space of functions (even and odd) of argument $x_i$. The value of the multiple is
\begin{align}
\label{Dynamical-Casimir-Value}
q_{x_i}=\mu_{x_i}^2+\alpha_{x_i}-3/4.
\end{align}
It follows from the above considerations that the eigenstates of each one-dimensional singular Dunkl oscillator span the space of a direct sum of two irreducible $\mathfrak{su}(1,1)$ representations; one for each parity case. The representation theory of $\mathfrak{su}(1,1)$ can be used to obtain the spectrum of $\mathcal{H}_{x_i}$. In point of fact, it is known \cite{Humphreys-1972} that in the positive discrete series of irreducible unitary representations of $\mathfrak{su}(1,1)$ in which the Casimir operator takes the value $\mathcal{C}=\nu(\nu-1)$, where $\nu$ is a positive real number, the spectrum of $J_0$ is of the form $n+\nu$, where $n$ is a non-negative integer. Given the identification \eqref{Identification-Algebra} and the Casimir values \eqref{Even}, \eqref{Odd} it follows that the spectrum of $\mathcal{H}_{x_i}$ is 
\begin{align}
\label{Energy-1D}
E_n^{+}=2n+\nu_{x_i}^{+}+1/2,\qquad E_{n}^{-}=2n+\nu_{x_i}^{-}+3/2,
\end{align} where 
\begin{align*}
\nu_{x_i}^{\pm}=2k_{x_i}^{\pm}+\mu_{x_i},
\end{align*}
and where $n$ is a non-negative integer. The $\pm$ sign is associated to the eigenvalues of the reflection $R_{x_i}$. The following conditions must hold on the values of the parameters:
\begin{align}
\label{Conditions}
\nu_{x_i}^{+}+1/2>0,\quad \text{and}\quad \nu_{x_i}^{-}+3/2>0.
\end{align}
It follows from \eqref{Energy-1D} that the spectrum of the full Hamiltonian \eqref{Dunkl-Singular-Osc} splits in four sectors labeled by the eigenvalues $(s_x,s_y)$ of the reflection operators $R_{x}$, $R_{y}$. The expression for the spectrum is
\begin{align}
\label{Full-Spectrum}
E_{n_xn_y}^{s_{x}s_{y}}=2(n_x+n_y)+\nu_{x}^{s_{x}}+\nu_{y}^{s_y}+\theta_{s_x}+\theta_{s_y}+1,
\end{align}
where $s_{x_i}=\pm 1$ and where 
\small
\begin{align}
\label{THETA}
\theta_{s_{x}}=
\begin{cases}
0 & \text{if $s_{x}=1$},\\
1 & \text{if $s_{x}=-1$}.
\end{cases}
\end{align}
\normalsize
It is understood that for example when $s_{x}=-1$, one should read $\nu_{x}^{s_{x}}$ as $\nu_{x}^{-}$.
\subsection{Exact solutions and separation of variables}
It is possible to obtain explicitly the wavefunctions $\Psi$ satisfying the Schr\"odinger equation
\begin{align}
\label{SCH}
\mathcal{H}\Psi=E\Psi,
\end{align}
associated to the Hamiltonian \eqref{Dunkl-Singular-Osc} in both Cartesian and polar coordinates.

\subsubsection{Cartesian coordinates}\hfill
\bigskip

\noindent
The Hamiltonian \eqref{Dunkl-Singular-Osc} obviously separates in Cartesian coordinates and in these coordinates the separated wavefunctions $\psi(x_i)$ are those of the one-dimensional singular Dunkl oscillator \eqref{Singular-Dunkl-1D}. The eigenfunctions $\psi(x)$ of $\mathcal{H}_{x}$ are easily seen to satisfy the differential equation
\small
\begin{align}
\label{Diff-Eq}
\psi''(x)+\frac{2\mu_{x}}{x}\,\psi'(x)+\left\{2E-x^2-\frac{\alpha_x+\mu_x}{x^2}\right\}\psi(x)+\left\{\frac{\mu_x-\beta_{x}}{x^2}\right\}R_{x}\psi(x)=0.
\end{align}
\normalsize
Since the reflection $R_{x}$ commutes with the one-dimensional Hamiltonian $\mathcal{H}_{x}$, the eigenfunctions can be chosen to have a definite parity. For the even sector, defined by $R_x\psi^{+}(x)=\psi^{+}(x)$, one finds that the normalizable solution to \eqref{Diff-Eq} is given by
\small
\begin{align}
\label{Plus-Sector}
\psi^{+}_{n_x}(x)=(-1)^{n_x}\sqrt{\frac{n_x!}{\Gamma(n_x+\nu_{x}^{+}+1/2)}}\,e^{-x^2/2}x^{2k_x^+}L_{n_x}^{(\nu_{x}^{+}-1/2)}(x^2),
\end{align}
\normalsize
with 
\begin{align*}
E_{n_x}^{+}=2n_x+\nu_x^{+}+1/2,
\end{align*}
and where $L_{n}^{(\alpha)}(x)$ are the Laguerre polynomials \cite{Koekoek-2010}. For the odd sector, defined by $R_{x}\psi^{-}(x)=-\psi^{-}(x)$, the wavefunctions are of the form
\small
\begin{align}
\label{Minus-Sector}
\psi^{-}_{n_x}(x)=(-1)^{n_x}\sqrt{\frac{n_x!}{\Gamma(n_x+\nu_x^{-}+3/2)}}\,e^{-x^2/2}x^{2k_x^-+1}L_{n_x}^{(\nu_x^{-}+1/2)}(x^2),
\end{align}
\normalsize
with 
\begin{align*}
E_{n_x}^{-}=2n_x+\nu_x^{-}+3/2.
\end{align*}
Hence, as announced, the wavefunctions of the two-dimensional Hamiltonian \eqref{Dunkl-Singular-Osc} split in four parity sectors labeled by the eigenvalues of the reflection operators $R_{x}$, $R_{y}$ and are given by
\begin{align}
\label{Full-Wave}
\Psi_{n_xn_y}^{s_xs_y}(x,y)=\psi_{n_x}^{s_x}(x)\psi_{n_y}^{s_y}(y),
\end{align}
with energies $E_{n_x,n_y}^{s_x,s_y}$ as in \eqref{Full-Spectrum} and with $\psi_{n_y}^{s_x}$ given by \eqref{Plus-Sector}, \eqref{Minus-Sector}. Using the orthogonality relation of the Laguerre polynomials, it is directly checked that the wavefunctions \eqref{Full-Wave} enjoy the orthogonality relation
\begin{align}
\label{Ortho}
\int_{-\infty}^{\infty}\int_{-\infty}^{\infty} \Psi_{n_xn_y}^{s_xs_y}(x,y)[\Psi_{n_x'n_y'}^{s_x's_y'}(x,y)]^{*}|x|^{2\mu_x}|y|^{2\mu_y}\,\mathrm{d}x\,\mathrm{d}y=\delta_{n_xn_x'}\delta_{n_yn_y'}\delta_{s_xs_x'}\delta_{s_ys_y'}.
\end{align}
Let us point out that a direct computation shows \cite{Genest-2013-1} that the Dunkl derivative \eqref{Dunkl-D} is anti-Hermitian with respect to the scalar product
\begin{align*}
\braket{f}{g}=\int_{-\infty}^{\infty}g(x)f^{*}(x)|x|^{2\mu_x}\,\mathrm{d}x.
\end{align*}
\subsubsection{Spacing of energy levels in the singular Dunkl oscillator}\hfill
\bigskip

\noindent
It is directly seen from \eqref{Energy-1D} that for generic values of $k_{x}^{+}$, $k_{x}^{-}$, the full spectrum of the one-dimensional singular Dunkl oscillator which comprises both the even and odd sectors is not equidistant, in contradistinction with the situation for the ordinary singular oscillator. An equidistant spectrum is obtained by taking $k_{x}^{+}=k_{x}^{-}=k_x$. In this case, the energies \eqref{Energy-1D} and wavefunctions can both be synthesized in single formulas which are close to the corresponding ones for the ordinary Dunkl oscillator \cite{Genest-2013-1}. In this case, one finds for the energies
\begin{align*}
E_{n_x}=n_{x}+\nu_x+1/2,\quad n_x=0,1,\ldots.
\end{align*}
The wavefunctions are expressed as
\begin{align*}
\psi_{n_x}(x)=e^{-x^2/2}x^{2k_x}H_{n_x}^{\nu_x}(x),
\end{align*}
where $H_{n}^{\gamma}(x)$ are the generalized Hermite polynomials \cite{Chihara-1978}
\begin{align}
\label{Generalized-H}
H_{2m+p}^{\gamma}(x)=(-1)^{m}\sqrt{\frac{m!}{\Gamma(2m+p+\gamma+1/2)}}\,x^{p}L_{m}^{(\gamma+p-1/2)}(x^2),
\end{align}
with $p\in\{0,1\}$. The wavefunctions of the full two-dimensional model have then the expression
\begin{align*}
\Psi_{n_x,n_y}(x,y)=e^{-(x^2+y^2)/2}x^{2k_x}y^{2k_y}H_{n_x}^{\nu_x}(x)H_{n_y}^{\nu_y}(y),
\end{align*}
with $E_{n_x,n_y}=(n_x+n_y)+\nu_x+\nu_y+1$, where $n_x$, $n_y$ are non-negative integers, as the corresponding energies. It is directly seen that upon taking $k_x=k_y=0$ in the above formulas, one recovers the results found in \cite{Genest-2013-1} for the Dunkl oscillator model.
\subsubsection{Polar coordinates}\hfill
\bigskip

\noindent
In polar coordinates
\begin{align*}
x=\rho \cos\phi,\quad y=\rho\sin\phi,
\end{align*}
the reflection operators have the action
\begin{align*}
R_{x}\Psi(\rho,\phi)=\Psi(\rho,\pi-\phi),\quad R_{y}\Psi(\rho,\phi)=\Psi(\rho,-\phi).
\end{align*}
The Schr\"odinger equation \eqref{SCH} associated to the Hamiltonian \eqref{Dunkl-Singular-Osc} takes the form
\begin{align}
\label{SCH-2}
\left\{\mathcal{A}_{\rho}+\frac{1}{\rho^2}\mathcal{B}_{\phi}\right\}\Psi(\rho,\phi)=E\Psi(\rho,\phi),
\end{align}
where $\mathcal{A}_{\rho}$ has the expression
\begin{gather*}
\mathcal{A}_{\rho}=-\frac{1}{2}\left\{\pd_{\rho}^2+\frac{1}{\rho}\pd_{\rho}\right\}-\frac{1}{\rho}(\mu_x+\mu_y)\pd_{\rho}+\frac{1}{2}\rho^2,
\end{gather*}
and where $B_{\phi}$ is given by
\begin{align*}
\mathcal{B}_{\phi}&=-\frac{1}{2}\pd_{\phi}^2+(\mu_x\,\mathrm{tan}\phi-\mu_y\,\mathrm{cot}\phi)\pd_{\phi}+\left\{\frac{\mu_x+\alpha_{x}}{2\cos^2\phi}\right\}+\left\{\frac{\mu_y+\alpha_{y}}{2\sin^2\phi}\right\}\\
&+\left\{\frac{\beta_x-\mu_x}{2\cos^2\phi}\right\}R_{x}+\left\{\frac{\beta_y-\mu_y}{2\sin^2\phi}\right\}R_{y}.
\end{align*}
\normalsize
It is easy to see that the equation \eqref{SCH-2} admits separation in polar coordinates. Upon taking $\Psi(\rho,\phi)=P(\rho)\Phi(\phi)$, we obtain the pair of ordinary differential equations
\begin{subequations}
\begin{gather}
\label{first}
\mathcal{B}_{\phi}\Phi(\phi)-\frac{m^2}{2}\Phi(\phi)=0,\\
\label{second}
\mathcal{A}_{\rho}P(\rho)+\left(\frac{m^2}{2\rho^2}-E\right)P(\rho)=0,
\end{gather}
\end{subequations}
where $m^2/2$ is the separation constant. The solutions to \eqref{first} split in four parity sectors labeled by the eigenvalues $s_{x}$, $s_{y}$ of the reflection operators $R_{x}$, $R_{y}$. The angular wavefunctions are found to be
\begin{align*}
\Phi^{s_xs_y}_{n}(\phi)=N_{n}\cos^{2k_x^{s_{x}}+\theta_{s_x}}\phi\;\sin^{2k_y^{s_{y}}+\theta_{s_y}}\phi\;P_{n-\theta_{s_x}/2-\theta_{s_y}/2}^{(\nu_y^{s_y}+\theta_{s_y}-1/2,\;\nu_x^{s_x}+\theta_{s_x}-1/2)}(\cos 2\phi),
\end{align*}
where $P_{n}^{(\alpha,\beta)}(x)$ are the Jacobi polynomials \cite{Koekoek-2010} and where $\theta_{s_x}$, $\theta_{s_y}$ are as in \eqref{THETA}. The admissible values of $n$ are as follows. If either $s_{x}$ or $s_{y}$ is negative, $n$ is a positive half-integer. If $s_{x}=s_{y}=1$, $n$ is a non-negative integer and if $s_{x}=s_{y}=-1$, $n$ is a positive integer. The normalization constant is
\begin{align*}
N_{n}=\sqrt{\frac{(2n+\nu_x^{s_x}+\nu_y^{s_y})\Gamma(n+\nu_x^{s_{x}}+\nu_y^{s_y}+\frac{\theta_{s_x}}{2}+\frac{\theta_{s_y}}{2})(n-\theta_{s_x}/2-\theta_{s_y}/2)!}{2\,\Gamma(n+\nu_x^{s_{x}}+\frac{\theta_{s_x}}{2}-\frac{\theta_{s_y}}{2}+1/2)\Gamma(n+\nu_{y}^{s_y}+\frac{\theta_{s_y}}{2}-\frac{\theta_{s_x}}{2}+1/2)}},
\end{align*}
where $\Gamma(x)$ is the Gamma function. The separation constant has the expression
\begin{align}
\label{Separation-Constant}
m^2=4(n+k_x^{s_{x}}+k_{y}^{s_{y}})(n+k_{x}^{s_{x}}+k_{y}^{s_{y}}+\mu_x+\mu_y),
\end{align}
and the wavefunctions obey the orthogonality relation
\begin{align*}
\int_{0}^{2\pi}\Phi^{s_xs_y}_{n}(\phi)\Phi^{s_x's_y'}_{n'}(\phi)|\cos \phi|^{2\mu_x}|\sin \phi|^{2\mu_y}\,\mathrm{d}\phi=\delta_{nn'}\delta_{s_x,s_x'}\delta_{s_y,s_y'}.
\end{align*}
The normalizable solution to the radial equation \eqref{second} is found to be
\begin{align*}
P_{\ell}(\rho)=N_{\ell}\,e^{-\rho^2/2}\rho^{2n+2k_x^{s_{x}}+2k_{y}^{s_y}}L_{\ell}^{(2n+2k_{x}^{s_{x}}+2k_{y}^{s_{y}})}(\rho^2).
\end{align*}
and in terms of the quantum numbers associated to the polar basis, the energies of the Hamiltonian \eqref{Dunkl-Singular-Osc} are
\begin{align*}
E_{n\ell}=2(n+\ell)+\nu_{x}^{s_x}+\nu_{y}^{s_y}+1,
\end{align*}
where $\ell$ is a non-negative integer. The normalization factor is
\begin{align*}
N_{\ell}=\sqrt{\frac{2\ell !}{\Gamma(2n+2k_{x}^{s_{x}}+2k_{y}^{s_y}+\ell+1)}}.
\end{align*}
and the radial wavefunctions obey the orthogonality relation
\begin{align*}
\int_{0}^{\infty}P_{\ell}(\rho)P_{\ell'}(\rho)\rho^{2\mu_x+2\mu_y+1}\,\mathrm{d}\rho=\delta_{\ell\ell'}.
\end{align*}
\subsection{Integrals of motion and symmetry algebra}
The integrals of motion of the two-dimensional Dunkl oscillator are most naturally obtained by combining the $\mathfrak{su}(1,1)$ dynamical operators of the one-dimensional model. We define
\begin{align}
\label{Ladder-Basis}
B_0=\mathcal{H}_{x}-\mathcal{H}_{y},\quad B_{+}=A_{x}^{\dagger}A_{y},\quad B_{-}=A_{x}A_{y}^{\dagger}.
\end{align}
It is directly checked that the operators \eqref{Ladder-Basis} are symmetries of the Hamiltonian \eqref{Dunkl-Singular-Osc}
\begin{align*}
[\mathcal{H},B_0]=[\mathcal{H},B_{\pm}]=0.
\end{align*}
A straightforward computation shows that the following commutation relations hold:
\begin{subequations}
\label{Symmetry-Algebra-1}
\begin{align}
[B_0,B_{\pm}]&=\pm 4 B_{\pm},\\
[B_{-},B_{+}]&=B_0^3+u_1B_0+u_2,
\end{align}
\end{subequations}
where
\begin{align*}
u_1=\xi_x R_{x}+\xi_{y}R_{y}-\mathcal{H}^2-w_{x}-w_{y},\quad u_2=\mathcal{H}(\xi_{y}R_{y}-\xi_{x}R_{x}+w_{x}-w_{y}),
\end{align*}
and with
\begin{align*}
\xi_{x_i}=2(\mu_{x_i}-\beta_{x_i}),\quad w_{x_i}=2(\mu_{x_i}^2+\alpha_{x_i}-3/4).
\end{align*}
Since the full Hamiltonian $\mathcal{H}$ given by \eqref{Dunkl-Singular-Osc} and the reflections $R_{x}$, $R_y$ are central elements, they will act as multiples of the identity in any irreducible representation of \eqref{Symmetry-Algebra-1} and consequently the operators $u_1$, $u_2$ can be treated as ``structure constants''.  

It follows from the above considerations that the singular Dunkl oscillator \eqref{Dunkl-Singular-Osc} is superintegrable with a cubic symmetry algebra given by \eqref{Symmetry-Algebra-1}. The energies of the Hamiltonian \eqref{Dunkl-Singular-Osc} could be derived algebraically from the irreducible representations of the algebra \eqref{Symmetry-Algebra-1} \cite{Vinet-1995}. The basis of operators $\{B_0,B_{\pm}\}$ generating the symmetry algebra \eqref{Symmetry-Algebra-1} and defined by \eqref{Ladder-Basis} will be referred to as the ``ladder'' basis.
\subsection{Symmetries, separability and the Hahn algebra with involutions}
When considering a Hamiltonian that admits separation of variables in more than one coordinate system, an alternative approach to finding the symmetry generators consists in identifying the symmetries responsible for the separation of variables \cite{Miller-1977}. We shall consider this approach here and relate it to the ladder approach of the preceding subsection.

The symmetry associated to the separation in Cartesian coordinates has already been found and is obviously given by
\begin{align*}
K_1=B_0=\mathcal{H}_{x}-\mathcal{H}_{y}.
\end{align*}
When acting on the separated wavefunctions in Cartesian coordinates $\Psi(x,y)$ given by \eqref{Full-Wave}, this operator is diagonal with eigenvalues
\begin{align*}
\lambda=2(n_x-n_y)+\nu_{x}^{s_x}-\nu_{y}^{s_y}+\theta_{s_x}-\theta_{s_y},\qquad n_{x},\, n_{y}\in\{0,1,\ldots\}.
\end{align*}
The symmetry associated to the separation of variables in polar coordinates can be obtained by analogy with the standard singular oscillator case \cite{Zhedanov-1991}. We consider the operator
\begin{align*}
K_2=(x \mathcal{D}_{y}^{\mu_y}-y \mathcal{D}_{x}^{\mu_{x}})^{2}-\frac{y^2}{x^2}(\alpha_x+\beta_x R_{x})-\frac{x^2}{y^2}(\alpha_{y}+\beta_{y}R_{y})-1/2.
\end{align*}
It is directly checked that $[\mathcal{H},K_2]=0$ and that $K_2$ is hence a symmetry. The assertion that $K_2$ is the symmetry associated to the separation of variables in polar coordinates stems from the following expression for $K_2$:
\begin{align*}
K_2=-2\mathcal{B}_{\phi}+(\alpha_{x}+\beta_{x}R_{x})+(\alpha_{y}+\beta_{y}R_{y})-2\mu_x\mu_y(1-R_{x}R_{y})-1/2,
\end{align*}
which is easily obtained by a direct computation. Thus the operator $K_2$ acts in a diagonal fashion on the separated wavefunctions $\Psi(\rho,\phi)$ with eigenvalues
\begin{align*}
\lambda=-m^2+\alpha_x+\alpha_{y}+\beta_x s_{x}+\beta_{y}s_{y}-2\mu_x\mu_y(1-s_{x}s_{y})-1/2,
\end{align*}
where $m^2$ is as given by \eqref{Separation-Constant}. The symmetry $K_2$ can be expressed in terms of the operators of the ladder basis \eqref{Ladder-Basis}. Upon inspection, one finds
\begin{align}
\label{Identification}
K_2=B_{+}+B_{-}+\frac{1}{2}B_0^2-\frac{1}{2}\mathcal{H}^2+\mu_x R_x+\mu_y R_{y}+2\mu_x\mu_y R_{x}R_{y}.
\end{align}
With this identification, the symmetry algebra \eqref{Symmetry-Algebra-1} can be written in terms of the symmetries $K_1$, $K_2$ and their commutator $K_3=[K_1,K_2]$. Using the commutation relations \eqref{Symmetry-Algebra-1}, the symmetry algebra becomes
\begin{subequations}
\label{Hahn-Algebra}
\begin{align}
[K_1,K_2]&=K_3,\\
[K_2,K_3]&=8\{K_1,K_2\}+\gamma_1K_1+\gamma_2,\\
[K_3,K_1]&=8K_1^2-16K_2+\gamma_3,
\end{align}
\end{subequations}
where the ``structure constants'' have the form
\begin{align*}
\gamma_1&=-8(2\beta_x R_{x}+2\beta_{y}R_{y}+4\mu_x\mu_y R_{x}R_{y}+w_{x}+w_{y}),\\
\gamma_2&=8\mathcal{H}(\xi_{y}R_{y}-\xi_{x}R_{x}+w_{x}-w_{y}),\\
\gamma_3&=16(\mu_x R_{x}+\mu_yR_y+2\mu_x\mu_y R_xR_y)-8\mathcal{H}^2.
\end{align*}
Under the transformation \eqref{Identification}, the symmetry algebra \eqref{Symmetry-Algebra-1} has become quadratic. The algebra \eqref{Hahn-Algebra} is a special case of the Askey-Wilson algebra $AW(3)$ known as the Hahn algebra \cite{Zhedanov-1991-2} with additional central involutions $R_x$, $R_{y}$; other presentations of the this algebra (without reflections) are found in \cite{Miller-2013}. The algebra \eqref{Hahn-Algebra} has the Casimir operator
\begin{align*}
\mathcal{Q}=8\{K_1^2,K_2\}+(56+\gamma_1)K_1^2-16K_2^2+K_3^2+2\gamma_2K_1+(2\gamma_3+16)K_2,
\end{align*}
which commutes with all generators $K_1$, $K_2$ and $K_3$. In the present realization, one finds that the operator $\mathcal{Q}$ takes the value
\begin{align*}
\mathcal{Q}=\zeta_1 \mathcal{H}^2-\zeta_2R_x-\zeta_3 R_{y}+\zeta_4 R_{x}R_{y}+\zeta_5,
\end{align*}
with 
\small
\begin{gather*}
\zeta_1=16\Big\{(\mu_x^2+\mu_y^2+\alpha_x+\alpha_y+2)+(\beta_x-2\mu_x)R_{x}+(\beta_y-2\mu_y )R_{y}-2\mu_x\mu_y R_{x}R_{y}\Big\},\\
\zeta_2=64\Big\{(\beta_x-2\mu_x)(\mu_y^2+\alpha_y-3/4)-\mu_x+\mu_x\alpha_y\Big\},\\ \zeta_3=64\Big\{(\beta_y-2\mu_y)(\mu_x^2+\alpha_x-3/4)-\mu_y+\mu_y\alpha_x\Big\},\\
\zeta_4=64(\beta_{x}\mu_{y}+\beta_{y}\mu_x-\beta_{x}\beta_{y}),\\
\zeta_5=64\Big\{\mu_x^2+\mu_y^2-\alpha_x(\mu_y^2+\alpha_y-3/4)-\alpha_y(\mu_x^2-3/4)-1/2\Big\}.
\end{gather*}
\normalsize
Because of the direct connection between the irreducible representations of the Askey-Wilson algebra $AW(3)$ and the Askey scheme of orthogonal polynomials \cite{Zhedanov-1991-2}, the occurrence of the Hahn algebra with reflections \eqref{Hahn-Algebra} as a symmetry algebra of the 2D singular Dunkl oscillator model suffices to establish that the dual Hahn polynomials act as overlap coefficients between the polar and Cartesian bases \cite{Zhedanov-1991}. This result contrasts with the situation in the case of the 2D Dunkl oscillator model, for which the overlap coefficients were found in terms of the dual $-1$ Hahn polynomials. This difference is explained by the fact that in the ordinary Dunkl oscillator case, the reflections anticommute with the raising/lowering operators and consequently the space of degenerate eigenfunctions of a given energy is labeled by the eigenvalues of the \emph{product} $R_{x}R_{y}$ and thus for example the sectors corresponding to $s_x=s_y=1$ and $s_x=s_y=-1$ are ``coupled''. In the singular oscillator case the spaces corresponding to different values of $s_x$, $s_y$ are fully ``decoupled''.
\section{The $2:1$ anisotropic Dunkl oscillator}
We shall now introduce our second two-dimensional Dunkl oscillator model: the 2D anisotropic Dunkl oscillator with a $2:1$ frequency ratio. The standard 2:1 oscillator is known to be one of the two-dimensional models which is superintegrable of order two and admits separations in both Cartesian and parabolic coordinates \cite{Winter-1965}; it is correspondingly of interest to consider its Dunkl analogue. It will be shown that this system is also second-order superintegrable, but does not seem to admit separation of variables except in Cartesian coordinates. We shall however present special cases of the general model for which separation in parabolic coordinates occurs.
\subsection{Hamiltonian, dynamical symmetries and spectrum}
The $2:1$ anisotropic Dunkl oscillator is defined by the Hamiltonian
\begin{align}
\label{Dunkl-Anisotropic}
\mathcal{H}=-\frac{1}{2}[(\mathcal{D}_{x}^{\mu_x})^2+(\mathcal{D}_{y}^{\mu_y})^2]+\frac{1}{2}(4x^2+y^2)+\frac{\alpha_{y}+\beta_{y}R_{y}}{2y^2},
\end{align}
where 
\begin{align*}
\alpha_{y}&=2k_{y}^{+}(k_{y}^{+}+\mu_y-1/2)+2k_{y}^{-}(k_{y}^{-}+\mu_y+1/2),\\
\beta_{y}&=2k_{y}^{+}(k_{y}^{+}+\mu_y-1/2)-2k_{y}^{-}(k_{y}^{-}+\mu_y+1/2),
\end{align*}
and with $k_y\in\mathbb{Z}$, $2k_y^{+}+\mu_{y}>-1/2$ and $2k_y^{-}+\mu_y>-3/2$. It is seen that \eqref{Dunkl-Anisotropic} corresponds to the combination of a one-dimensional singular Dunkl oscillator in the $y$ direction and an ordinary one-dimensional Dunkl oscillator with twice the frequency in the $x$ direction. The dynamical symmetries of the $y$ part of the anisotropic oscillator \eqref{Dunkl-Anisotropic} described by the Hamiltonian
\begin{align}
\label{Y-Part}
\mathcal{H}_{y}=-\frac{1}{2}(\mathcal{D}_{y}^{\mu_y})^2+\frac{1}{2}y^2+\frac{\alpha_{y}+\beta_{y}R_{y}}{2y^2},
\end{align}
have been studied in the previous section. The dynamical operators $A_{y}^{\dagger}$, $A_{y}$ are defined by \eqref{Dynamical-OPS} and together with $\mathcal{H}_{y}$ they generate the $\mathfrak{su}(1,1)$ Lie algebra \eqref{Singular-Dynamical-Algebra} with the invariant operator $Q_{y}$ defined \eqref{Dynamical-Casimir} taking the value \eqref{Dynamical-Casimir-Value}. The spectrum of $\mathcal{H}_{y}$ is known to be of the form
\begin{align*}
E_{n}^{+}=2n+\nu_{y}^{+}+1/2,\quad E_{n}^{-}=2n+\nu_{y}^{-}+3/2,
\end{align*}
where $n$ is a non-negative integer.
The dynamical symmetries of the $x$ part of the Hamiltonian \eqref{Dunkl-Anisotropic}
\begin{align}
\label{X-Part}
\mathcal{H}_{x}=-\frac{1}{2}(\mathcal{D}_{x}^{\mu_x})^2+2x^2,
\end{align}
are easily obtained. We introduce the operators
\begin{align}
\label{Anisotropic-Dynamical}
c_{x}=\sqrt{2}(x+\frac{1}{2}\mathcal{D}_{x}^{\mu_x}),\quad c_{x}^{\dagger}=\sqrt{2}(x-\frac{1}{2}\mathcal{D}_{x}^{\mu_{x}}).
\end{align}
It is directly checked that the following commutation relations hold
\begin{gather*}
[\mathcal{H}_{x},c_{x}]=-2c_{x},\quad [\mathcal{H}_{x},c_{x}^{\dagger}]=2c_{x}^{\dagger},\quad [c_{x},c_{x}^{\dagger}]=2+4\mu_x R_{x},\quad \{c_x,c_{x}^{\dagger}\}=2\mathcal{H}_{x},\\
[\mathcal{H}_x,R_{x}]=0,\quad \{c_{x},R_{x}\}=0,\quad \{c_{x}^{\dagger},R_{x}\}=0.
\end{gather*}
The dynamical algebra \eqref{Anisotropic-Dynamical} is directly identified with the $sl_{-1}(2)$ algebra \cite{Zhedanov-2011}. The algebra \eqref{Anisotropic-Dynamical} admits the Casimir operator
\begin{align*}
Q=c_{x}^{\dagger}c_{x}R_{x}-\mathcal{H}_{x}R_{x}+R_{x},
\end{align*}
which commutes with all the dynamical operators $\mathcal{H}_{x}$, $c_{x}$, $c_{x}^{\dagger}$ and acts as a multiple of the identity:
\begin{align*}
Q=q\mathbb{I},\quad q=-2\mu_x.
\end{align*}
Using the representation theory of $sl_{-1}(2)$ \cite{Zhedanov-2011}, the expression for the spectrum of $\mathcal{H}_{x}$ is found to be
\begin{align*}
E_{n}=2n+2\mu_x+1,\quad n=0,1,\ldots
\end{align*}
It follows that the spectrum of the two-dimensional anisotropic Dunkl oscillator \eqref{Dunkl-Anisotropic} is given by
\begin{align}
\label{Energy-Anisotropic}
E^{s_{y}}_{n_x,n_y}=2(n_{x}+n_{y})+2\mu_x+\nu_{y}^{s_{y}}+\theta_{s_{y}}+3/2,
\end{align}
where $\nu_{y}^{\pm}=2k_{y}^{\pm}+\mu_y$.
\subsection{Exact solutions and separation of variables}
It is possible to write down in Cartesian coordinates the exact solutions of the Schr\"odinger equation corresponding to the Hamiltonian \eqref{Dunkl-Anisotropic}. The wavefunctions are again of the form $\Psi(x,y)=\varphi(x)\psi(y)$ where $\varphi(x)$ is a wavefunction of the ordinary Dunkl oscillator with frequency $2$ and $\psi(y)$ is a wavefunction of the singular Dunkl oscillator.

The solutions to the equation $\mathcal{H}_{x}\varphi(x)=E\varphi(x)$ have been derived in \cite{Genest-2013-1,Rozen-1994}. They take the form
\begin{align*}
\varphi_{n_x}(x)=2^{(\mu_x+1/2)/2}\,e^{-x^2}H_{n_x}^{\mu_x}(\sqrt{2}x),
\end{align*}
where $H_{n}^{\mu}(x)$ denotes the generalized Hermite polynomials defined in \eqref{Generalized-H}. The corresponding energies are
\begin{align*}
E_{n_x}=2n_{x}+2\mu_x+1,\quad n_{x}=0,1,\ldots,
\end{align*}
The solutions to the equation $\mathcal{H}_{y}\psi(y)=E\psi(y)$ have been found in the preceding section in terms of Laguerre polynomials and are given by \eqref{Plus-Sector} and \eqref{Minus-Sector}. It follows that the exact solutions of the Schr\"odinger equation of the  $2:1$ anisotropic Dunkl oscillator are of the form
\small
\begin{align*}
\Psi_{n_xn_y}^{s_{y}}(x,y)=\sqrt{\frac{2^{\mu_x+1/2}\,n_y !}{\Gamma(n_{y}+\nu_{y}^{s_y}+\theta_{y}+1/2)}}e^{-(2x^2+y^2)/2}y^{2k_{y}^{s_y}+\theta_{s_y}}H_{n_x}^{\mu_x}(\sqrt{2}x)L_{n_y}^{(\nu_{y}^{s_y}+\theta_{y}-1/2)}(y^2),
\end{align*}
\normalsize
with energies given by \eqref{Energy-Anisotropic} and where $s_{y}=\pm 1$.

A direct inspection of the Hamiltonian \eqref{Dunkl-Anisotropic} shows that this Hamiltonian does not seem to admit separation of variable in any other coordinate system. This situation differs with that of the standard anisotropic oscillator in the plane \eqref{Standard-Anisotropic} which admits separation of variable in parabolic coordinates.
\subsection{Integrals of motion and symmetry algebra}
The dynamical operators of the anisotropic oscillator \eqref{Dunkl-Anisotropic} can again be used to obtain its symmetry generators and establish the superintegrability of the model. Proceeding as in the Section 2, we introduce the operators
\begin{align*}
F_0=\mathcal{H}_{x}-\mathcal{H}_{y},\quad F_{+}=c_{x}^{\dagger}A_{y},\quad F_{-}=c_{x}A_{y}^{\dagger},
\end{align*}
where $\mathcal{H}_x$ is given by \eqref{X-Part}, $\mathcal{H}_y$ by \eqref{Y-Part}, $A_{y}$, $A_{y}^{\dagger}$ by \eqref{Dynamical-OPS} and $c_{x}$, $c_{x}^{\dagger}$ by \eqref{Anisotropic-Dynamical}. A direct examination shows that the operators $F_0$ and $F_{\pm}$ are symmetries of the anisotropic Dunkl oscillator Hamiltonian \eqref{Dunkl-Anisotropic}; $[\mathcal{H}_{x},F_0]=[\mathcal{H}_{x},F_{\pm}]=0$. A straightforward computation shows that these operators generate the following quadratic algebra:
\begin{subequations}
\label{Algebra-2}
\begin{gather}
[F_0,F_{\pm}]=\pm 4 F_{\pm},\quad \{F_{\pm},R_{x}\}=0,\quad [F_0,R_{x}]=0,\\
[F_{-},F_{+}]=\frac{3}{2}F_0^2+z_1 F_0^2R_{x}+z_2 F_0+z_3 F_0R_{x}+z_4R_{x}+z_5.
\end{gather}
\end{subequations}
with
\begin{gather*}
z_1=\mu_x,\quad z_2=-\mathcal{H},\quad z_3=-2\mu_x\mathcal{H},\\
z_4=\mu_x\mathcal{H}^2+2\mu_x(\xi_{y}R_{y}-w_{y}),\quad z_5=(\xi_{y}R_{y}-w_{y})-\frac{1}{2}\mathcal{H}^2.
\end{gather*}
The operators $\mathcal{H}$ and $R_{y}$ are central elements in the algebra \eqref{Algebra-2}. Introducing the operator
\begin{align*}
F_1=\frac{1}{\sqrt{2}}(F_{+}+F_{-}),
\end{align*}
one finds that
\begin{align}
\label{Dunkl-Runge-Lenz}
F_1=\frac{1}{2}\{\mathcal{D}_{y}^{\mu_{y}},(x\mathcal{D}_{y}^{\mu_{y}}-y\mathcal{D}_{x}^{\mu_{x}})\}+xy^2-\frac{x}{y^2}(\alpha_{y}+\beta_{y}R_{y}),
\end{align}
where $\{x,y\}=xy+yx$. The operator $F_1$ is thus seen to be a Dunkl analogue of the generalized Runge-Lenz vector \cite{Goldstein-1980}, which is the symmetry associated to the separation of variables in parabolic coordinates in the standard anisotropic oscillator case. The Dunkl Hamiltonian \eqref{Dunkl-Anisotropic} does not however separate in this coordinate system due to the presence of singular terms in both the $x$ and $y$ parts.

We shall now examine two special cases of the $2:1$ anisotropic Dunkl oscillator for which the separation in parabolic coordinates is however possible. These special cases are obtained by removing the reflections in the $x$-part \eqref{X-Part} which prevents the separation in parabolic coordinates.
\subsection{Special case I}
The first special case of the $2:1$ anisotropic Dunkl oscillator that we consider is described by the Hamiltonian
\begin{align}
\label{Special-Case-I}
\mathcal{H}_{I}=-\frac{1}{2}[\pd_{x}^2+(\mathcal{D}_{y}^{\mu_y})^2]+\frac{1}{2}(4x^2+y^2)+\frac{\alpha_y+\beta_{y}R_{y}}{2y^2},
\end{align}
with the usual quantization conditions
\begin{gather*}
\alpha_{y}=2k_{y}^{+}(k_{y}^++\mu_y-1/2)+2k_{y}^{-}(k_y^{-}+\mu_y+1/2),\\
\beta_{y}=2k_{y}^{+}(k_{y}^++\mu_y-1/2)-2k_{y}^{-}(k_y^{-}+\mu_y+1/2),
\end{gather*}
where $k_{y}\in\mathbb{Z}$. It is easily seen that this Hamiltonian is obtained from \eqref{Dunkl-Anisotropic} by taking $\mu_x=0$ and hence it is superintegrable and its symmetry algebra is obtained directly from \eqref{Algebra-2}. Moreover, the Hamiltonian \eqref{Special-Case-I} corresponds to the combination of a standard harmonic oscillator in the $x$ direction with a singular Dunkl oscillator in the $y$ direction. In parabolic coordinates
\begin{align*}
x=\frac{1}{2}(u^2-v^2),\qquad y=uv,
\end{align*}
the Schr\"odinger equation $\mathcal{H}_{I}\Psi(u,v)=E\Psi(u,v)$, with $\mathcal{H}_{I}$ the Hamiltonian given by \eqref{Special-Case-I}, takes the form
\begin{align*}
\left[\mathcal{C}_{u}+\mathcal{C}_{v}\right]\Psi(u,v)=-2E(u^2+v^2)\Psi(u,v),
\end{align*}
where
\begin{align*}
\mathcal{C}_{u}=\pd_{u}^2+\frac{2\mu_y}{u}\pd_{u}-\frac{\mu_y}{u^2}(1-R_y)-\frac{1}{u^2}(\alpha_y+\beta_y R_{y})-u^6.
\end{align*}
Since the reflection operator $R_{y}$ commutes with $\mathcal{H}_{I}$, the wavefunction $\Psi(u,v)$ can be taken to have a definite parity. In the even sector, defined by the relation $R_y\Psi^{+}(u,v)=\Psi^{+}(u,v)$, the separation Ansatz $\Psi^{+}(u,v)=U(u)V(v)$ yields the following pair of ordinary differential equations:
\begin{subequations}
\begin{align}
\label{first-1}
U''(u)+\frac{2\mu_{y}}{u}U'(u)+\left\{2Eu^2-\frac{\alpha_{y}+\beta_{y}}{u^2}-u^6\right\}U(u)&=\Lambda^{+} U(u),\\
\label{second-1}
V''(v)+\frac{2\mu_{y}}{v}V'(v)+\left\{2Ev^2-\frac{\alpha_{y}+\beta_{y}}{v^2}-v^6\right\}V(v)&=-\Lambda^{+} V(v),
\end{align}
\end{subequations}
where $\Lambda^{+}$ is the separation constant. Upon imposing the condition $U(-u)=U(u)$, the solution to the equation \eqref{first-1} is seen to be given by
\begin{align*}
U^{+}(u)=e^{-u^4/4}u^{2k_y^{+}}\,B(\nu_{y}^{+}-1/2;0;E;\frac{\Lambda^{+}}{\sqrt{2}};\frac{u^2}{\sqrt{2}}),
\end{align*}
where $B(\alpha;\beta;\gamma;\delta;z)$ is the Heun biconfluent function. The Heun biconfluent function is defined as the solution to the differential equation
\begin{align*}
B''(z)-\frac{(-\alpha+\beta z+2z^2-1)}{z}B'(z)-\frac{1}{2}\frac{(\alpha\beta+\beta+\delta+z(2\alpha+4-2\gamma))}{z}B(z)=0,
\end{align*}
with initial conditions $B(0)=1$ and $B'(0)=(\alpha\beta+\beta+\delta)/(2\alpha+2)$. Similarly, the solution for $V(v)$ is directly given by
\begin{align*}
V^{+}(v)=e^{-v^4/4}v^{2k_y^{+}}\,B(\nu_{y}^{+}-1/2;0;E;\frac{-\Lambda^{+}}{\sqrt{2}};\frac{v^2}{\sqrt{2}}).
\end{align*}
In the odd sector, defined by the relation $R_{y}\Psi(u,v)=-\Psi(u,v)$, we obtain
\begin{align*}
U^{-}(u)=e^{-u^4/4}u^{2k_y^{-}+1}\,B(\nu_{y}^{-}+1/2;0;E;\frac{\Lambda^{-}}{\sqrt{2}};\frac{u^2}{\sqrt{2}}),
\end{align*}
and
\begin{align*}
V^{-}(v)=e^{-v^4/4}u^{2k_y^{-}+1}\,B(\nu_{y}^{-}+1/2;0;E;\frac{-\Lambda^{-}}{\sqrt{2}};\frac{v^2}{\sqrt{2}}).
\end{align*}
In parabolic coordinates, the operator $F_{1}$ given in \eqref{Dunkl-Runge-Lenz} can readily be shown to be diagonal and its eigenvalues can be related to the separation constants $\Lambda^{\pm}$. As is the case for the standard anisotropic oscillator \eqref{Standard-Anisotropic}, the parameters $\Lambda$ obey a transcendental equation and cannot be expressed explicitly \cite{Vinet-1995}.
\subsection{Special case II}
Another special case of the $2:1$ anisotropic Dunkl oscillator which admits separation of variables in parabolic coordinates is described by the Hamiltonian
\begin{align}
\label{Special-Case-II}
\mathcal{H}_{II}=-\frac{1}{2}[\pd_{x}^2+(\mathcal{D}_{y}^{\mu_y})^2]+\frac{1}{2}(4x^2+y^2).
\end{align}
The Hamiltonian \eqref{Special-Case-II} is obtained by taking $\mu_x=0$, $\alpha_{y}=0$ and $\beta_{y}=0$ in \eqref{Dunkl-Anisotropic}. It corresponds to the combination of an ordinary Dunkl oscillator in the $y$ direction and a standard oscillator with twice the frequency in the $x$ direction. For this Hamiltonian, the symmetry algebra is not a special case of the algebra \eqref{Algebra-2}; this follows from the fact that the dynamical operators of the singular Dunkl oscillator involve the squares of the dynamical symmetries of the ordinary Dunkl oscillator.

The dynamical operators in the present special case are rather the creation/annihilation operators of the standard oscillator
\begin{align*}
g_{x}=\sqrt{2}(x+\frac{1}{2}\pd_{x}),\quad g_{x}^{\dagger}=\sqrt{2}(x-\frac{1}{2}\pd_{x}),
\end{align*}
which are part of the algebra
\begin{align*}
[g_{x},g_{x}^{\dagger}]=2,\quad [\mathcal{H}_{x},g_{x}]=-2g_{x},\quad [\mathcal{H}_{x},g_{x}^{\dagger}]=2g_{x}^{\dagger},
\end{align*}
and those of the one-dimensional Dunkl (or parabosonic) oscillator
\begin{align*}
h_{y}=\frac{1}{\sqrt{2}}(y+\mathcal{D}_{y}^{\mu_y}),\quad h_{y}^{\dagger}=\frac{1}{\sqrt{2}}(y-\mathcal{D}_{y}^{\mu_{y}}),
\end{align*}
which obey the $sl_{-1}(2)$ algebra relations
\begin{align*}
[h_{y},h_{y}^{\dagger}]=1+2\mu_{y}R_{y},\quad [\mathcal{H}_{y},h_{y}]=-h_{y},\quad [\mathcal{H}_{y},h_{y}^{\dagger}]=h_{y}^{\dagger}.
\end{align*}
The symmetries of the Hamiltonian \eqref{Special-Case-II} are of the form
\begin{align*}
T_0=\mathcal{H}_{x}-\mathcal{H}_{y},\quad T_{+}=g_{x}^{\dagger}h_{y},\quad T_{-}=g_{x}h_{y}^{\dagger},
\end{align*}
and they generate the following algebra:
\begin{gather*}
[T_0,T_{\pm}]=\pm 3T_{\pm},\quad \{T_{\pm},R_{y}\}=0,\quad [T_0,R_{y}]=0,\\
[T_{+},T_{-}]=\frac{3}{2}T_0+\mu_yT_0R_{y}+\mathcal{H}(\mu_{y}R_{y}-1/2)+2\mu_y R_{y}.
\end{gather*}
Similarly to the preceding special case, the separation of variable in parabolic coordinates can be performed and the wavefunctions in these coordinates satisfy the biconfluent Heun equation. Since the computations are analogous to those already presented here, we omit the details.
\section{Conclusion}
In this paper, we have considered two extensions of the Dunkl oscillator model: one with additional singular terms in the potential and the other with singular terms and a $2:1$ frequency ratio. We showed that the singular Dunkl oscillator is second-order superintegrable and exhibited its symmetry generators. We also identified the symmetry algebra as an extension with central involution operators of a special case of the Askey-Wilson algebra $AW(3)$. We also obtained the exact solutions in both Cartesian and polar coordinates. For the $2:1$ anisotropic Dunkl oscillator, we showed the system to be also second-order superintegrable, exhibited the symmetries and obtained the algebra they generate. Special cases for which separation of variables in parabolic coordinates occurs were also considered and their symmetry algebras found. In one instance, it was shown that the wavefunctions in parabolic coordinates obey the biconfluent Heun equation.

The models investigated here can be considered as generalizations of the standard singular oscillator \eqref{Standard-Singular} and $2:1$ anisotropic oscillator \eqref{Standard-Anisotropic} with the derivatives replaced by the Dunkl derivatives. In this context, it is natural to consider the Hamiltonian corresponding to the Dunkl-Coulomb problem
\begin{align*}
H=-\frac{1}{2}\nabla^{2}_{\mathcal{D}}+\frac{\alpha}{r},
\end{align*}
where $\nabla^2_{\mathcal{D}}$ is the Dunkl-Laplacian, which in Cartesian coordinates reads
\begin{align*}
\nabla^{2}_{\mathcal{D}}=(\mathcal{D}_{x}^{\mu_x})^2+(\mathcal{D}_{y}^{\mu_y})^2.
\end{align*}
The spectrum of this Hamiltonian can be evaluated algebraically using the observation that $r$, $r\nabla^2_{\mathcal{D}}$ and $x\pd_{x}+y\pd_{y}$ obey the $\mathfrak{sl}_{2}$ relations \cite{deBie-2011}. The expression for the values of the energy then depend on a single quantum number and the spectrum of $H$ exhibits accidental degeneracies; this suggests that the Dunkl-Coulomb Hamiltonian in the plane is also superintegrable. In the standard case, it is known that the singular oscillator model \eqref{Standard-Singular} can be related to the Coulomb problem via the Levi-Civita mapping. This is not so with the singular Dunkl oscillator and Dunkl-Coulomb problem since the Levi-Civita mapping amounts to a passage from Cartesian to parabolic coordinates, a coordinate system in which the Dunkl Laplacian does not separate. Hence the Dunkl-Coulomb problem is genuinely different from the singular Dunkl oscillator and shall be considered elsewhere.

It would be of interest in a future study to identify and characterize other novel superintegrable systems of Dunkl type. The Dunkl-type models defined on the circle, in the 3-dimensional Euclidean space and on the 2-sphere are of particular interest. Given the relation of Dunkl oscillator models and $-1$ polynomials, this study could provide further insight into the emerging Bannai-Ito scheme of $-1$ orthogonal polynomials.

\section*{Acknowledgements}
V.X.G. holds an Alexander-Graham-Bell scholarship from the Natural Sciences and Engineering Research Council of Canada (NSERC) and from Fonds de Recherche du Qu\'ebec-Nature et Technologies. The research of L.V. is supported in part by NSERC.
\section*{References}

\end{document}